%% file: CensusTracts_Globecom16.tex
\documentclass[conference]{IEEEtran}

\usepackage[utf8]{inputenc}
\usepackage{color}
\usepackage{cite}
\usepackage{graphicx}
\usepackage{epsfig}
\usepackage{epstopdf}
\usepackage{psfrag}
\usepackage{mathrsfs}
\usepackage{acronym}
\usepackage{amsfonts}
\usepackage{amsmath}
\usepackage{amssymb}
\usepackage{verbatim}
\usepackage{amsthm}
\usepackage{caption}
\usepackage{subfig}
\usepackage{url}
\usepackage{float}
\begin{document}

\def\mb{\mathbf}
\def\mc{\mathcal}
\def\mbb{\mathbb}
\def\e #1{\mathbb{E}\left\lbrace #1 \right\rbrace}
\def\hl #1{\textcolor{red}{#1}}

\title{Census Tract License Areas: Disincentive for Sharing the 3.5GHz band?}

 \author{\IEEEauthorblockN{Elma Avdic,~Irene Macaluso,~Nicola Marchetti~and~Linda Doyle}
 \IEEEauthorblockA{CONNECT, Research Centre for Future Networks and Communications\\University of Dublin, Trinity College, Ireland\\
 Email: \{avdice,~macalusi,~marchetn,~linda.doyle\}@tcd.ie}
 }

\maketitle

\begin{abstract}
 	
Flexible licensing model is a necessary enabler of the technical and procedural complexities of Spectrum Access System (SAS)-based sharing framework. The purpose of this study is to explore the effectiveness of 3.5GHz Licensing Framework - based on census tracts as area units, areas whose main characteristic is population. As such, the boundary of census tract does not follow the edge of wireless network coverage. We demonstrate why census tracts are not suitable for small cell networks licensing, by (1) gathering and analysing the official census data, (2) exploring the boundaries of census tracts which are in the shape of non-convex polygons and (3) giving a measure of effectiveness of the licensing scheme through metrics of area loss and the number of people per census tract with access to spectrum. Results show that census tracts severely impact the effectiveness of the licensing framework since almost entire strategically important cities in the U.S. will not avail from spectrum use in 3.5GHz band. 

Our paper does not seek to challenge the core notion of geographic licensing concept, but seeks a corrective that addresses the way the license is issued for a certain area of operation. The effects that inappropriate size of the license has on spectrum assignments lead to spectrum being simply wasted in geography, time and frequency or not being assigned in a fair manner. The corrective is necessary since the main goal of promoting innovative sharing in 3.5 GHz band is to put spectrum to more efficient use.
\end{abstract}


\input{Introduction}
\input{Sharing_Framework}

\input{Census_Tracts_Study}

\input{Discussion_and_Results}

\section*{Conclusions}
\label{sect:Conclusions}
The main conclusion from the detailed analysis of census tracts is that they do not provide a good basis on which to build a licensing framework. The FCC rightly recognizes that smaller license areas provide the potential for unlocking efficient sharing of spectrum in 3.5GHz band using small cell deployments. But as our results have shown, even under propagation conditions that are favorable, many unit license areas are rendered unusable and that potential is not unlocked.


The problem of spectrum waste over the license area, reflected through the percentage of area loss and number of people precluded from spectrum use points to the fact that the boundaries of census tracts are not appropriate boundaries for a license area in a geographic licensing scheme envisioned for 3.5 GHz band. Not only that they mismatch with the spectrum propagation of small cell deployments, the main deployments that  could put 3.5 GHz spectrum to more efficient use, but also finding the CBRS-allowed remains a problem with no solution even if the boundary limits are less stringent. In highly dense urban areas, census tracts are so small that it is not possible to form the CBRS-allowed area at all to fit the diverse spectrum usage so the interference between adjacent tracts is reduced. As we have shown, the smaller the census tracts are, more spectrum is wasted over the area. 



Getting the licensing framework right is essential. The landscape in spectrum management has changed, in that it is accepted that some form of spectrum sharing will be part of the future. However it is still an open question as to which approaches will succeed. The 3.5GHz SAS-based sharing promises much and may be hampered in delivering if census tracts are at the core of its licensing framework. 

Therefore we are pointing out that it is worth re-considering the licensing scheme, because the size and the shape of the license area matters. We propose the following geographic licensing models to take into consideration towards building an effective licensing framework that can fully support the SAS potential. Geographic area of deployment calculated based on the predicted actual spectrum usage should be the base area unit of the licensing scheme, i.e. area whose size is not fixed nor created for purposes other than spectrum usage. Describing the actual spectrum usage of an operator's network is a research question we plan to address in our future work, in order to investigate how different network deployments need to be so that they can be allocated in the same license area and even in the same spectrum chunk. 

Furthermore, the currently existing geographic licensing models for spectrum sharing are implemented on the borders of the U.S. with Canada and Mexico in 800MHz band. These models are built on the basis of geographic sharing while giving assurance that spectrum is assigned in a fair manner to adjacent operators, whose edge users will get the service under these conditions. Defined sharing and protection zones reduce the need for grey space non-allowed area and there is always someone using the frequencies in time and space, showing in a way that the operator networks can overlap and still avail from spectrum use. 

\section*{Acknowledgement}
This work is supported by the Seventh Framework Programme for Research of the European Commission under grant ADEL-619647 and the Science Foundation Ireland under grant CONNECT 13/RC/2077. We would like to thank Dr. Tim Forde for his insightful comments and valuable feedback which improved the quality of this paper.
%

\end{document}

%% file: Introduction.tex
\section{Introduction}
\label{sect:intro}
The licensing framework for sharing the 3.5 GHz band is based on the notion of \textit{census tracts}. Census tracts are geographical areas defined on the basis of population statistics with their area boundaries not expected to change much over time. The Federal Communications Commission has adopted the census tract demographic areas as the  \textit{licensing area units} - their cartographic boundaries  will serve as boundaries of allowed licensed operation in the 3.5GHz band\cite{RO}. In this paper, we explore whether this choice makes sense and what implications this choice has for efficient spectrum usage. 

Attention from the academia and spectrum regulation sector in the past few years has been directed towards the Spectrum Access System (SAS) - a technologically sophisticated system of multiple databases that will manage the tiers of diverse users, set in a complex spectrum sharing environment of the 3.5GHz band.  Specially three areas of SAS-based sharing have been discussed in the literature: (1) architectural implementation and feasibility of the SAS system \cite{SASarch}, \cite{GoogleSAS} , (2) regulatory and policy issues around license design, spectrum rights and enforcement classes for the future SAS system \cite{LEHR}, \cite{WEISS}, and (3) incumbent tier coexistence with small cell operators in the band \cite{Incumbents}, \cite{Radars}. 
%

\begin{figure}
 \centering
 \includegraphics[width=9cm]{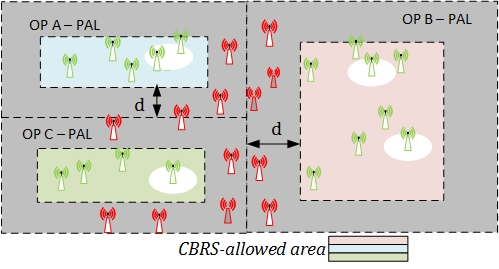}
 \caption{Adjacent census tracts allocated to different operators: the set-back distance and the sterile, grey space area in which the red nodes cannot be used.}
 \label{fig:censustracts}
 \end{figure}

%
Despite being the basic unit on which the licensing framework for 3.5 GHz band is based, the impact of census tracts on the effectiveness of the framework has not been analyzed in detail.
A \textit{license} to use spectrum is granted with the purpose of avoiding harmful interference that the licensee may experience from the neighboring spectrum users. Therefore to efficiently manage the spectrum use,  spectral and geographical neighbors have to be issued with compatible licenses. For the 3.5 GHz band, the spectrum users are Citizens Broadband Radio Service (CBRS) users, a new class of service established by the FCC in order to promote innovative sharing for 3.5 GHz band \cite{NPRM}. The\textit{ rights to transmit} (specified in the license content) for CBRS users will define the area of operation, the frequencies, and the power levels of operation. One license for CBRS users will be issued per census tract. Therefore to limit the risk of interference to neighbouring census tracts, CBRS nodes cannot be deployed too close to the boundary of the census tracts. More specifically, if licenses for two adjacent census tracts are issued to different CBRS networks - the nodes of these networks can only be deployed within the \textit{set-back distance} from the border of the census tract. The distance is determined by the propagation characteristics of the band and FCC technical rules on CBRS operation \cite{FNPRM}. As illustrated in Fig.\ref{fig:censustracts}, under these conditions, the existence of a grey space area where CBRS networks cannot be deployed points to the problem of \textit{spectrum waste over the area}, which we address in this paper.


 
Google Inc. has raised the problem of the census tract within the stakeholders' filings to the FCC\footnote[1]{Stakeholders filings to the Commission are available at ECFS home page: http://apps.fcc.gov/ecfs/).}, under the 3.5 GHz docket 12-354 \cite{Comments_RF}. To substantiate their concerns, Google carried out a fairly simplified geometry analysis of the census tract area boundaries in which they are represented as equivalent circular areas. Within these circles, the set-back area in which no CBRS devices can be deployed was shown to be substantial.


In the same filing, other stakeholders expressed their support for the Commission's proposal on census tracts. The arguments in favor included: (1) population characteristics of census tracts will give operators an option to plan their network deployments to target certain groups of customers for the service they aim to provide in that area and (2) the official census database could be incorporated into SAS to more effectively support the accuracy of spectrum usage information, i.e. geographic data about spectrum users. 
 \label{sect:introduction}

The purpose of our paper is to conduct a wider analysis of census tracts and further explore such geographic licensing scheme. It is not clear whether the existence of intersecting\footnote[2]{Intersection of census tracts is illustrated in \cite{Comments_RF} as a location where several census tracts meet, e.g. Washington Convention Center. If an operator wishes to serve the users at this location, it would need to aggregate several licenses for all census tracts that meet there. Details about the rules in Licensing Framework for 3.5 GHz are elaborated in Section \ref{sect:framework}.} census tracts or tracts of inappropriate size across the country  would severely jeopardize the flexibility of the licensing framework for 3.5 GHz band or whether it is an issue only for a number of specific locations. 
Therefore we use official census data (cartographic files and census statistics) in order to analyze boundaries of real census tracts. We conduct studies on entire cities, classified in the U.S. census databases as Major Economic Areas (MEAs)\footnote[3]{http://transition.fcc.gov/oet/info/maps/areas/}. Sample cities for the study are: New York City (Manhattan Burough), Washington DC and  San Francisco, CA - selected as urban area types based on their diverse population densities.

As a proxy of spectrum utilisation, we use two efficiency metrics to analyze the effectiveness of census tracts-based licensing. 
The first metric of Area Loss Percentage (ALP) as introduced in \cite{Comments_RF}, takes into account the boundaries of the license areas defined through the ratio of CBRS-allowed area and the corresponding census tract area.
The second takes account of the essential characteristic of a census tract, namely the demographics. Hence we introduce a metric we term Population of Census Tract with Access to Spectrum (PCTAS), which gives us an opportunity to evaluate spectrum utilisation in terms of real consumers of network capacity who will be affected by the sharing rules for 3.5 GHz band. \\
Our results show that even in the best propagation conditions in the band, for which the area that would be off-limits to CBRS nodes is minimised - the implications can be stark. For example, in the case of Manhattan, 82\% of census tracts license areas will not be available for deployment if one license is issued per one census tract. For the less favourable propagation conditions, the Manhattan area cannot be used for spectrum sharing in 3.5 GHz at all, and results for Washington DC and San Francisco show that less that 10\% of census tracts in both cities is actually available for deployments.

%
To begin with the framework itself, in Section \ref{sect:framework} we introduce the aspects of SAS-based sharing and potential issues with the licensing scheme. New classes of users created for this band are explained with the specific technical rules that will affect the licensing framework whose area unit is a census tract and we overview the details of census tracts. In Section \ref{sect:study} we describe the study conducted followed by the results presentation and discussion in Section \ref{sec:results}. The last section concludes the paper.

%% file: Sharing_Framework.tex
\section{Sharing Framework and SAS}
\label{sect:framework} 
\begin{figure*}
\centering
\includegraphics[scale=0.3]{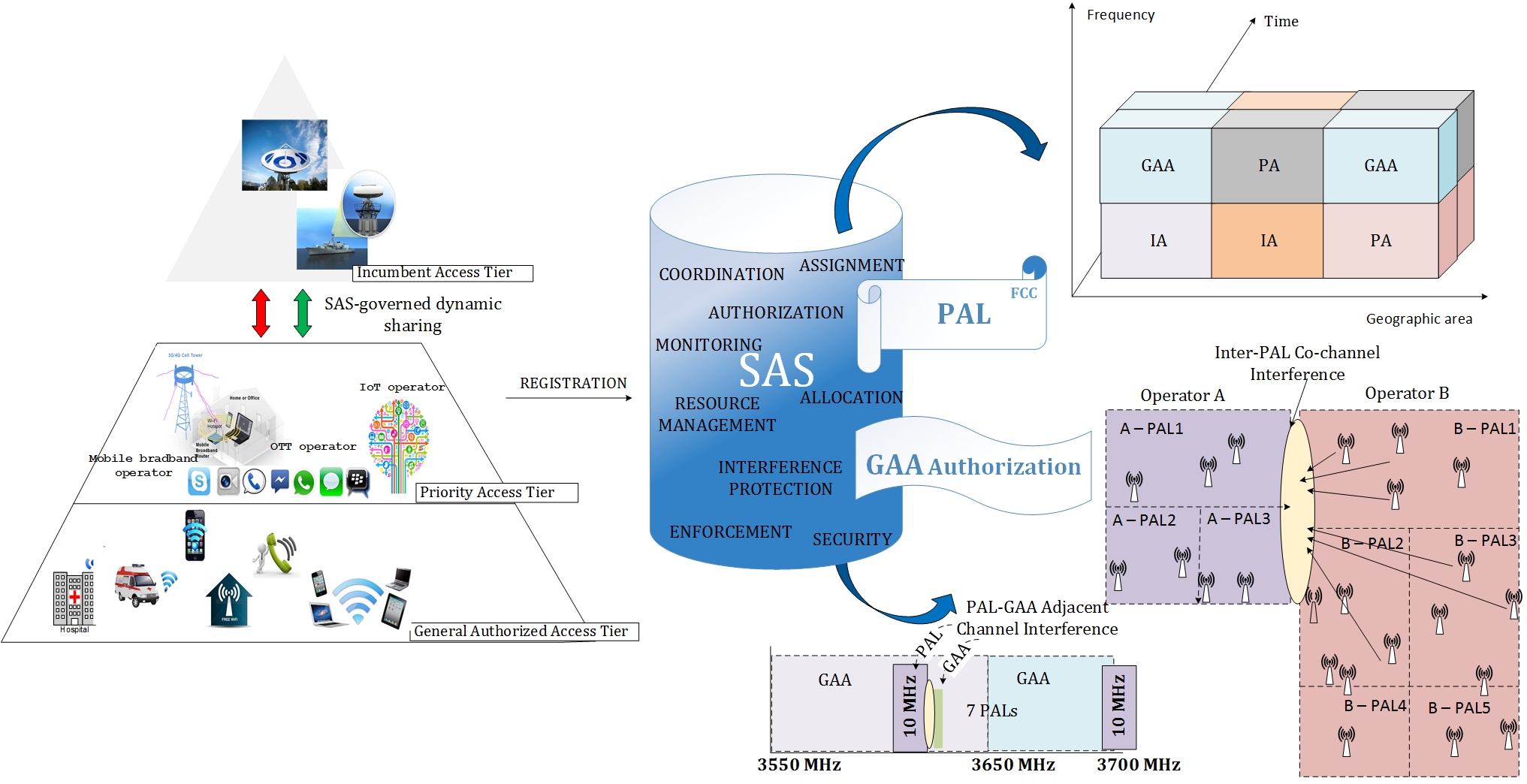}
\caption{Tiers of SAS and SAS functionalities: The 3.5GHz band becomes open to licensed and opportunistic use, accommodating critical operations facilities (hospitals, public safety networks), business and residential users alongside with federal users and fixed satellite stations which encumber the band.}
\label{fig:SAS}
\end{figure*}
The President’s Council of Advisors on Science and Technology (PCAST) Report \cite{PCAST} has called for innovation in spectrum sharing by mapping the technology and regulatory enablers and advancing them both so that sharing becomes a norm. The sharing framework for 3.5 GHz band accommodates diversity of users, envisions a range of different technologies to be used and supports many different applications to enable efficient sharing.

%

%


One of the many inspiring messages of the PCAST report was that wireless users will get to share military spectrum. To embed this promising sharing diversity into the set of effective and clear regulatory rules of operation, the FCC has put a stamp of approval on such message by first - creating an entirely new class of service for the band \cite{NPRM}, and second - adopting the technical rules of CBRS operation \cite{FNPRM}. The broad range of future usage is suggested in the title itself, Citizens Broadband Radio Service. The type of usage and the classes of services under the CBRS name, as depicted in Fig.\ref{fig:SAS}, range from licensed carrier cells, fixed wireless broadband to advanced home networking and any other uses. 

Incumbent Access (IA) users, as primary holders of spectrum rights have the highest priority to access the spectrum which is reflected in exclusion and protection zones in which Priority Access (PA) and  General Authorized Access (GAA) users cannot operate. For the 3.5 GHz band, the IA users are federal users (military high-powered radars on ship platforms across the coastline) and grandfathered fixed satellite services (FSS). Federal incumbent use is a matter of national security and therefore such users need continuous protection from interference. However, national defence missions are usually executed in shorter times and in limited geographic areas, therefore unused portions of spectrum in space, time and frequency can be reallocated to other users. These users will receive protection from harmful interference that commercial users in lower tiers would generate; moreover they are not required to mitigate interference they generate to lower tiers. 

\subsection{The Licensing scheme and Frequency Assignments}
\textit{PA users} need to be issued with a license to access spectrum through a geographic licensing scheme - to operate within the boundaries of one census tract area. 
The U.S. is divided into 74,000 census tracts, whose boundaries follow political (e.g., subdivision of counties) or geographic boundaries like rivers and roads. They are constructed to encompass a population of 2,000-4,000 on average.

A PA network operator will be issued a license to operate within one census tract in one 10 MHz channel. The license duration is 3 years and is non-renewable. However, aggregation of spectrum is allowed in all three dimensions(space, frequency and time). Spatially, one PA user is allowed to aggregate an unlimited number of census tract areas in order to serve more users. In frequency, a maximum of four 10 MHz channels within one census tract area can be aggregated. In time, PA license (PAL) applicants can apply for 2 consecutive license terms within the first applicant window, gaining in such way the license for 6 years. 

The size of the census tracts varies, from areas less then a square mile in high dense urban regions, to 85,000 square miles in less inhabited rural regions. They often intersect roads, strategic touristic attractions or institutions of high commercial interest. Their intersection combined with the technical rules for a licensed operation means that an operator may have to acquire several adjacent census tracts licenses to deploy a small cell network and serve one particular building. On the other hand, when census tracts are too large, spectrum is wasted per area because of exclusivity of only one PAL operating in that tract per one 10MHz chunk. \\
\textit{GAA users} do not need to get a license and access spectrum opportunistically, similarly to an unlicensed mode of operation with the difference of being licensed-by-rule in the framework. This means that SAS assigns GAA users dynamically based on the demand, whenever and wherever spectrum is free from PA use. It also means that GAA devices have to be FCC technically certified (ability to tune to given frequencies or having embedded sensing capabilities in the device). \\
Despite the fixed bandwidth that SAS will assign to PA users (multiple of 10MHz), according to the demand and interference conditions SAS may move them to other 10 MHz spectrum chunks, if needed. To GAA users SAS will allocate from the pool of frequencies instead, i.e. any frequency free from PA use in the portion of 150 MHz, where interference constraints are satisfied. The dynamic assignments of 10MHz chunks to PAs and free frequencies from GAA pool of 150MHz, performed by a SAS system will follow \textit{the demand and supply} principle whilst providing assurance that some level of assignment convention is met. The directive on more efficient spectrum use also means that spectrum is assigned contiguously and continuously\footnote[4]{Initially, FCC proposed frequency assignment for GAA users of minimum 50 percent of the 150 MHz to be reserved for them in any census tract. After commenters questioned the proportional approach as a potential cause of uncertainty in the marketplace, FCC revised the proposal and concluded that a maximum of 70 MHz can be reserved for PALs in any given license area. This means that GAA users may access all of 150 MHz in areas where spectrum is free (no PALs issued or in use) or up to 80 MHz where all PALs are in use.}. For this reason, to assure fairness among users in the framework - GAA users are allowed in the enitre band of 150 MHz as long as incumbent and PA users tiers are not utilising the specific channels.

While the GAA users do not need licenses, since they get authorized and allocated by the SAS once they register, the concept of census tracts has implications on them. This is because the rules of operation adopt the same boundary limits for all CBSDs in the band, which means that neither PA nor GAA users can access the grey space strip area as shown in Fig.\ref{fig:censustracts}, in order to reduce the effect of interference leakage into adjacent census tracts.

 As depicted in Fig.\ref{fig:SAS}, the SAS is in charge of PA and GAA assignments and authorization, tiers coordination and management of interference due to tier-interactions. The types of interference that may occur in tier interactions are: (1) \textit{co-channel interference} (from PA and GAA to IA, between two adjacent PA users, between nearby GAAs, between PA and GAA users) and (2) \textit{adjacent channel interference} (from PA and GAA to IA, between PA users across adjacent license areas, between PA users within a license area, between PA and GAA users and between nearby GAAs). The SAS needs to coordinate the users in the band dynamically and potentially across multiple bands, so it also needs an embedded mechanism for monitoring and reporting spectrum usage to be able to manage the spectrum usage automatically.
 
%

Finally, CBRS devices (CBSDs) are divided into two groups, group A is small cell indoor/outdoor low power use and group B is point-to-point use (higher power devices)\footnote[5]{All CBSDs are required to register with a SAS and provide their location, antenna height above the ground, authorization status they request for (PA or GAA), FCC ID number, serial number of the device. Optionally, CBSDs should report on their sensing capabilities. Since frequency and power assignments are established by the SAS, all CBSDs must execute SAS instructions within 60 seconds.} Technical parameters around CBSD specifications can be found in \cite{RO} and we later summarise the specific parameters used in our study are in Table \ref{tab:study}.

%% file: Census_Tracts_Study.tex
\section{Census Tracts Study} 
\label{sect:study}

In this section, we describe the study conducted and the approach to the problem. First, we introduce the selection criteria for the census tract areas. Methods of collecting and analysing data for the study are described in brief, followed by the propagation analysis and the method for computing the \textit{CBRS-allowed area}, i.e. the portion of the area of a census tract in which CBSD deployment is allowed. To evaluate census tracts-based licensing scheme, we introduce two efficiency metrics, which take into account two main census tracts characteristics: area and their population. 

\subsection{Census tracts datasets}
\label{sect:data}

Traffic profile of metropolitan areas (particularly highly dense, urban areas) consists mostly of indoor residential and commercial traffic, but also outdoor WiFi traffic. The rules on technical operations in 3.5GHz band describe the PA and GAA spectrum usage corresponding to such traffic profiles, with the small cell deployments for indoor and outdoor categories of CBSDs. Based on their strategic importance and census statistics data, we selected 3 sample cities. The subjects of our study are: (1) New York City (Manhattan Borough), NY, (2) Washington, DC and (3) San Francisco, CA.  

The area of Manhattan, as an example of highly dense and urban area, is divided into 288 very small census tracts, with an extremely high population density of 69,467 people per square mile. The population density in urban area profiles of San Francisco and Washington DC is much lower (17,179 and 9,856 population per square mile, respectively), but census tracts areas are still small. 

%
%

The data for this study comes from the official U.S. census databases\footnote[6]{http://www.census.gov/geo/maps-data/index.html}. 
Census tracts are presented in the form of cartographic boundary files, given in .kml and .shp format\footnote[7]{We used .kml files, which are based on WGS84 geodetic system. World Geodetic system 1984 is the current geodetic system being used by GPS and U.S. DoD to satisfy mapping, charting and geodetic requirements. It is geocentric and globally consistent within 1m.}. To apply Cartesian geometry which is required by our study, we used the UTM (Universal Transverse Mercator) projection maps to convert geodetic coordinates of census tracts to Cartesian friendly, UTM coordinates (WGS84).
 
\subsection{Propagation Analysis}
\label{sec:propagation}
\begin{table}[!t]
\centering
\caption{Propagation Analysis - parameters}
\label{tab:study}
\begin{tabular}{|| c || c || c || c || c || p{5cm} |}
\hline
\bfseries Deployment & \bfseries Building Loss & \bfseries Path Loss & \bfseries Set-back \\ \hline
Outdoor & - & 110 dB & 2.1 km \\ \hline
Indoor residential & 10 dB (wood) & 100 dB & 663 m \\ \hline
Indoor commercial & 20 dB (metal) & 90 dB & 210 m \\ \hline
\end{tabular}

\end{table}



Being above the 3GHz threshold up to which mobile cellular spectrum usage is ideal, the 3.5 GHz\footnote[8]{Refering to the 3.5GHz band means the 150 MHz of spectrum between 3550-3700 MHz.} band is not ideal for exclusive, licensed and commercial mobile broadband usage. The core advantage of the band is a huge potential for geographic sharing. Having low powered deployments with the limited propagation range of the band will allow more users to operate in closer proximity\footnote[9]{The limited propagation of 3.5 GHz needs a technology that requires less range than macrocell networks to meet users demand. Propagation characteristics of 3.5GHz make the signal decay faster, which is why low-powered and dense small cell deployments empower this band. Small cells bring higher spatial and spectral reuse with its applications, since their larger density reduces the risk of interference in geography and spectrally which results in increased frequency reuse and network capacity. In addition, signal propagation of the band still allows flexible topologies, appropriate for non-line-of-site use.}. 

Geographic-based licensing in this band is on the level of a census tract. Therefore, the power limits on a license boundary are proposed by the FCC in the rules for technical operation in 3.5GHz \cite{RO}. The rules specify the conditions in which neighboring CBRS deployments should operate to reduce the risk of interference due to tier-interactions. 


 Anywhere along PA service area boundaries between different CBRS users, a signal strength level limit of -80 dBm, measured by a 0 dBi isotropic antenna in a 10 MHz bandwidth is proposed. The path loss amount required to meet the boundary limit is the difference of: allowed EIRP (30 dBm/10MHz), the boundary limit and building loss where existing, depending on the type of deployments. The distance, calculated from path loss free space model formula, for frequency of 3.6GHz (the frequency range is 3550-3700 MHz) and antenna gains of 0 dBi is the required set-back distance under different building losses. We summarise in Table \ref{tab:study} the path loss for different cases and the corresponding set-back distances.

In the following section, we use this distance to set the constraint on a boundary of the license area in order to compute the portion of the area that would be off-limits to CBSDs deployments, i.e. area in which CBSDs cannot meet the adopted boundary signal strength limit. As summarised in Table \ref{tab:study}, CBSD deployments considered are indoor residential and indoor commercial and we used full-power CBSDs, Category A.

\subsection{Computing the CBRS-allowed area}
\label{sec:model}
Census tracts polygons are \textit{non-convex polygons}, irregular in shape and in size. In general, they are layed out as sets of multiple polygons, each of them given as a set of points  in the geodetic coordinate system\footnote[10]{To obtain the list of points, coordinates conversion is applied so that the boundary constraint in optimisation model could be set for Euclidan distance (distance between geodetic coordinates is Haversine distance).}.

The problem of computing the \textit{CBRS-allowed area} is formulated as a collection of quadratic programming problems with quadratic constraints. The assumption underlying the model is that the adjacent census tracts are allocated to different operators. This case can be interpreted as a worst case scenario, but extremely important one to look at since one of the core ideas of 3.5GHz framework is to allow operators to operate locally. For each of the original polygons we compute the largest polygon (if exists) whose distance from the boundary of the original one is at least the set-back distance determined by propagation analysis, based on FCC technical requirements for CBRS devices.

Let us denote by $\{P_1,...,P_N\}$ the set of points in UTM that describe a polygon. For each point $P_i$, we determine a point $P_i^{*}$ such that the distance between  $P_i$ and $P_i^{*}$  is minimized, whilst a minimum distance $d$ (which is the set-back distance) between $P_i^{*}$  and every $P_j$ $\in\{P_1, P_2...,P_N\}$  is guaranteed, as formulated in (\ref{problem}). The set of points $\{P_1^{*}, P_2^{*}, ...,P_N^{*}\}$ defines the new polygon. 


\begin{equation}
\label{problem}
\begin{aligned}
& \underset{P_i^{*}}{\min} & & (P_i^{*}-P_i)^2\\
&\text{s.t.} & & (P_i^{*}-P_j)^2\geq d^2, \forall j \in \{1,...,N\}.
\end{aligned}
\end{equation}

It is important to note that in case the length of the segments that define the original polygon is greater than $2\times d$, the solution to (\ref{problem}) could include points in boundary of the original polygon. To avoid this problem, we 'densify' the original polygon by adding additional points any time the length of a segment is larger than $d$. This also improves accuracy in the proximity of corners.

\subsection{Metrics}
\label{sec:metrics}

Here, we introduce two efficiency  metrics to evaluate the census tract-based licensing. They are selected so that they encompass the main characteristics of census tracts: their boundary and their population. 
The area of each census tract polygon and the CBRS-allowed area\footnote[11]{As noted in \ref{sec:model}, CBRS-allowed area defines the region where CBRS nodes can be deployed, under the rules on boundary signal strength limits adopted by FCC for CBRS operations.} are computed in order to evaluate the licensing scheme through the efficiency metrics of \textit{Area Loss Percentage} and \textit{Population of Census Tract with Access to Spectrum}. 
The area of each polygon is computed as the area of an irregular polygon with known coordinates, based on Green's theorem.

The ALP metric, as defined in (\ref{eqn:arealoss})
 takes into account the ratio between the CBRS-allowed area, denoted as $A_{\rm CBRS}$, and the area of the original census tract polygon, denoted as $A_{\rm CT}$. Under the propagation analysis applied, ALP  gives the percentage of the area which cannot be used for sharing under technical rules for 3.5 GHz band.
 


\begin{equation}\label{eqn:arealoss}
\begin{aligned}
   &{\rm ALP}=1-\frac{A_{\rm CBRS}}{A_{\rm CT}}, 0<{\rm ALP}\leqslant 1.
\end{aligned}
\end{equation}

To account for the basic characteristic of census tracts, i.e. their population - we introduce the metric PCTAS, defined in (\ref{eqn:populationloss}) as the percentage of the area of a census tract in which spectrum is not wasted, weighted by the population count $P$ of the census tract.  

\begin{equation}\label{eqn:populationloss}
\begin{aligned}
    &{\rm PCTAS}=(1-{\rm ALP})\times P.
\end{aligned}
\end{equation}

Under the propagation analysis applied and under the assumption of uniformely distributed population, PCTAS gives the population in census tract that can consume network capacity under the 3.5GHz spectrum sharing rules in the cities selected for the study.







%% file: Discussion_and_Results.tex
\section{Results}
\label{sec:results}
\begin{figure*}
\centering
\subfloat[Case I, set-back distance of 210m ]{\includegraphics[width=6.5cm]{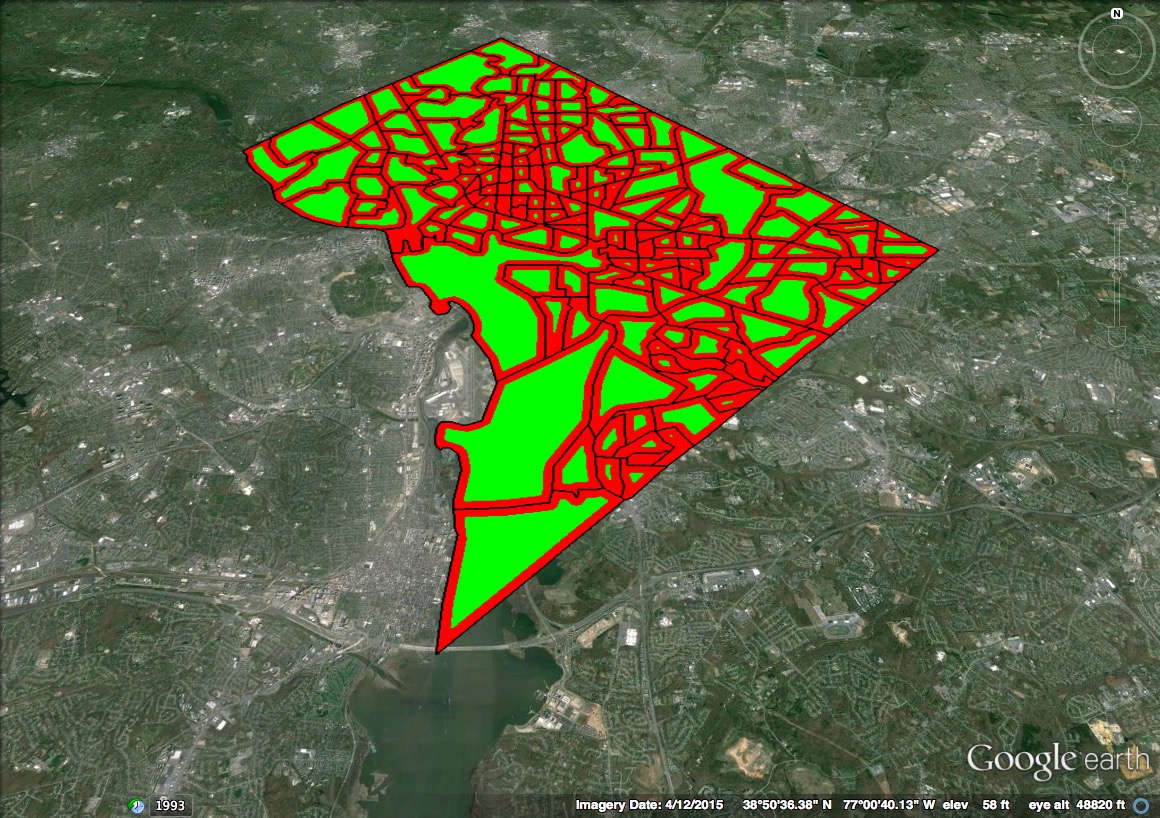}
\label{DCd1}}
\hfil
\subfloat[Case II, set-back distance of 663m]{\includegraphics[width=6.5cm]{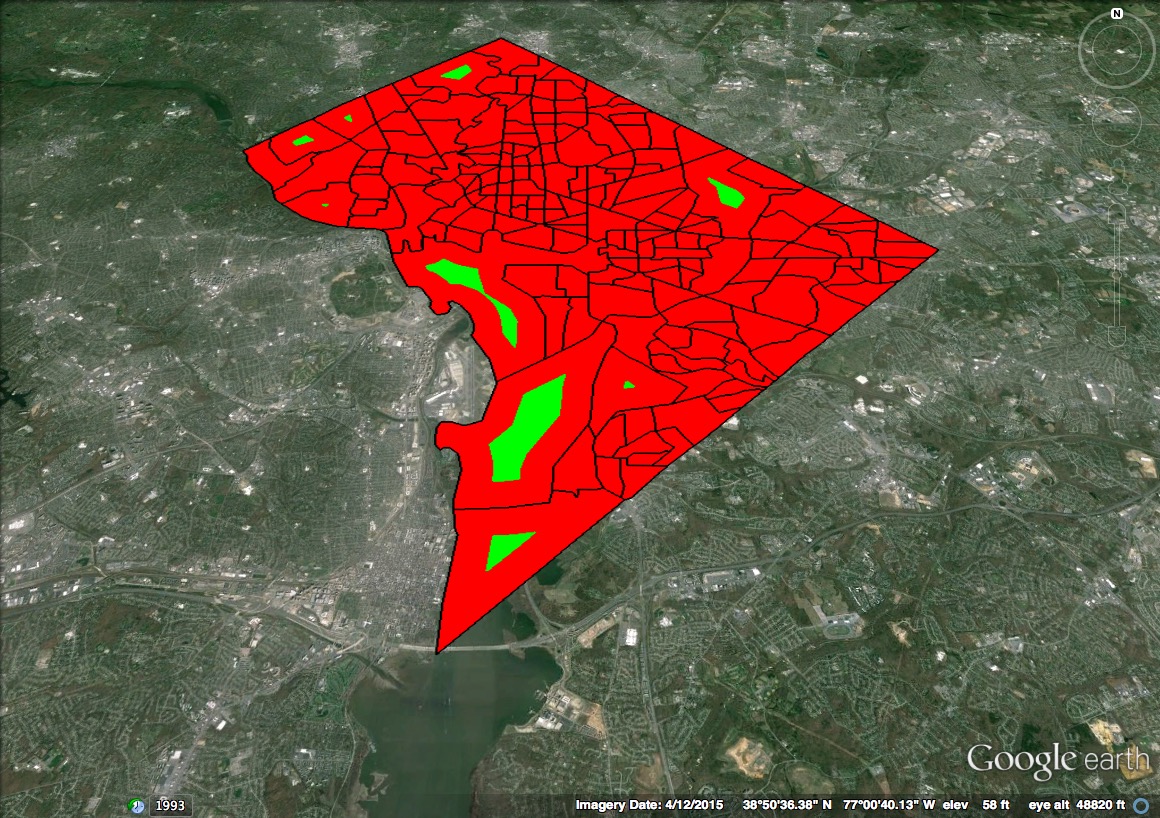}
\label{DCd2}}
\caption{Washington DC area}
\label{Washington}
\end{figure*}
\begin{figure*}
\centering
\subfloat[Case I, set-back distance of 210m ]{\includegraphics[width=8.5cm]{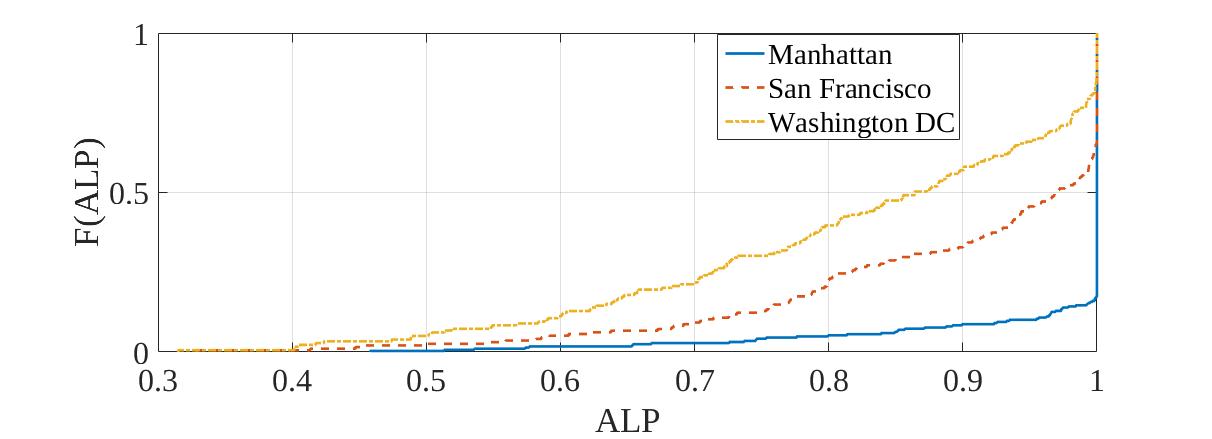}
\label{CDFAREALOSS}}
\hfil
\subfloat[Case II, set-back distance of 663m]{\includegraphics[width=8.5cm]{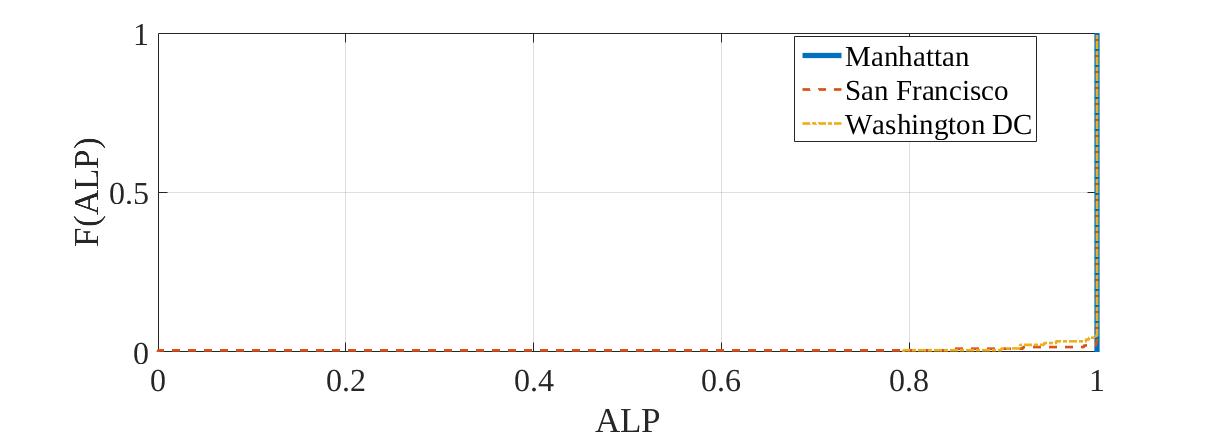}
\label{CDFAREALOSS2}}
\caption{CDF of ALP - Area Loss Percentage}
\label{CDF_AREALOSS2}
\end{figure*}
\begin{figure*}
\centering
\subfloat[Case I, set-back distance of 210m ]{\includegraphics[width=8.5cm]{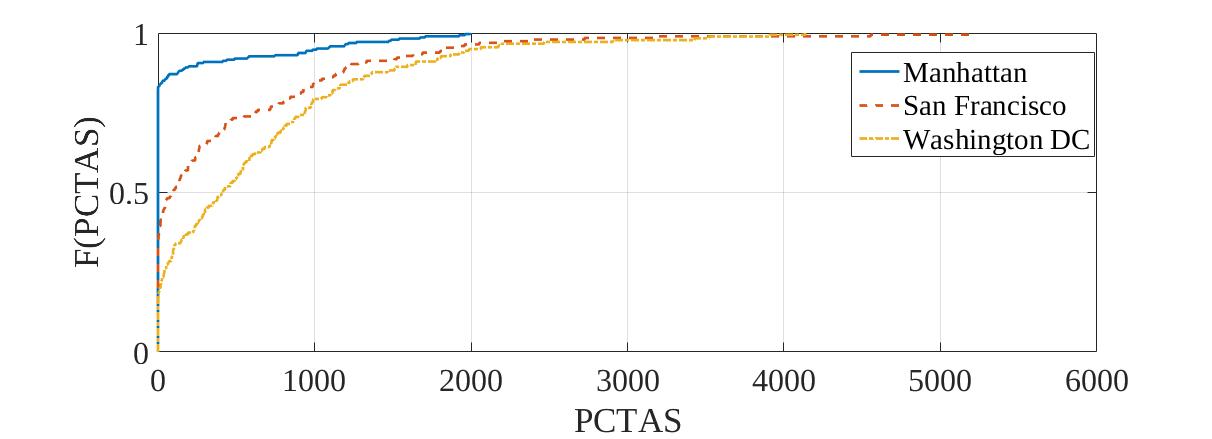}
\label{CDFPOPLOSS}}
\hfil
\subfloat[Case II, set-back distance of 663m]{\includegraphics[width=8.5cm]{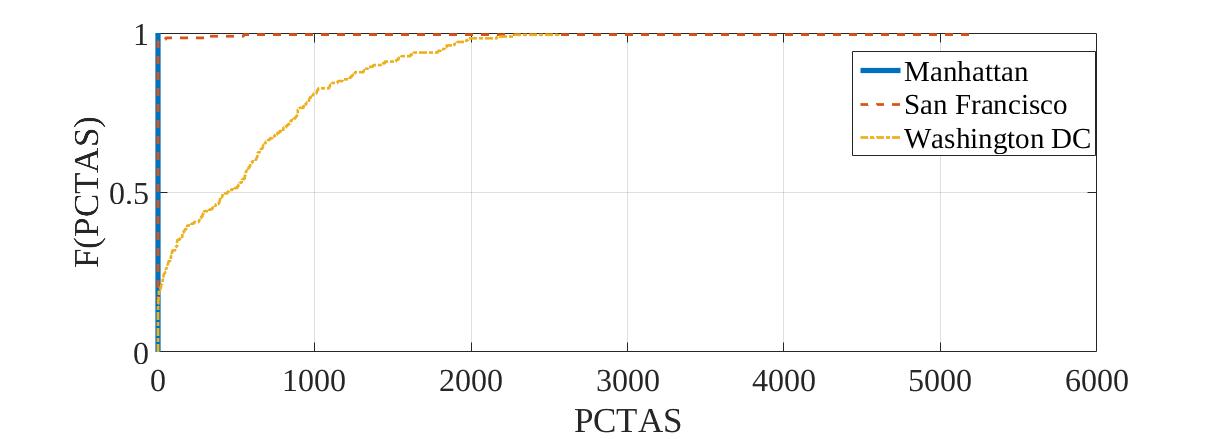}
\label{CDFPOPLOSS2}}
\caption{CDF of PCTAS - Population of Census Tract with Access to Spectrum}
\label{CDF_POPLOSS}
\end{figure*}

The results of our analysis are presented in Fig. \ref{Washington}, Fig. \ref{CDF_AREALOSS2} and Fig. \ref{CDF_POPLOSS}. To give a visual insight into the problem, Fig.\ref{DCd1} (210m set-back distance) and Fig.\ref{DCd2} (663m set-back distance) show the results of the area study for Washington DC. Green areas correspond to the portion of census tracts where CBRS deployment is allowed. Red areas correspond to areas off-limits to CBRS deployment. Compared to Manhattan and San Francisco, Washington DC results show a good outcome in terms of spectrum waste over the area for Washington. In fact, in Case I, the red areas cover almost entirely Manhattan and 35\% of San Francisco. Only 18\% of census tracts in  Manhattan could accommodate CBRS deployments to satisfy the boundary constraint on a minimum distance dictated by propagation analysis (i.e., the best propagation conditions case). The impact of 663m set-back distance is severe in all three cities. As seen on Fig.\ref{DCd2}, only 9 tracts, out of 179 tracts of the District of Columbia - will be available to accommodate CBRS deployments under these conditions. For 2.1km set-back distance, no census tract will be available for use in all the considered cities. \\
If we look at the cumulative distribution function (CDF) for the two efficiency metrics described in Section \ref{sec:metrics}, we can get an insight into spectrum utilisation for the cities analyzed. In the Case I, as shown in Fig.\ref{CDFAREALOSS}, 82\% of Manhattan census tracts will not be available for CBRS deployment. For the rest of the Manhattan areas, where deployment is possible, the area that is available for deployment is limited. For example, the probablity that ALP is less than 50\% is 0.0035. It means that even in the case the boundary limit can be satisfied, the potential for spectrum sharing could hardly be unlocked. In Case I, San Francisco area will be unavailable for deployments in about 35\% of census tracts, but the probablity that ALP is less than 50\% is still very small, namely 0.025. In the second case of a larger set-back distance, the Fig.\ref{CDFAREALOSS2} shows that Manhattan area will not be available for deployments at all, and only 3-4\% of Washington DC and San Francisco census tracts could be used as license areas.
\\
We now look at the unavailability of spectrum for network capacity consumers within census tract boundaries. Fig.\ref{CDF_POPLOSS} shows the CDF of PCTAS for 210m and 663m cases of propagation conditions. In the Case I, Fig.\ref{CDFPOPLOSS}, we can see that the probability that no one in Manhattan can consume the network service is 82\%. Further, it is certain that less than half of the average number of people per census tract in Manhattan (5500) can be served. Also, with probability $1$, less than three quarters of the average number of people per census tract (4087) in San Francisco will be served. Finally, in Washington DC the probability that less than average number of people per census tract (3360) will have a network service is 0.977. In Washington tracts with size of the population larger than 4000 people per tract, no one will take advantage of spectrum use. In the Case II, Fig.\ref{CDFPOPLOSS2}, no one in Manhattan can be served, almost no one in San Francisco and only less than 2000 people per census tract in Washington DC can consume network capacity.

%% file: CensusTracts_Globecom16.bbl
\begin{thebibliography}{15}

\bibitem{RO}
In the Matter of Amendment of the Commission's Rules with Regard to Commercial Operations in the 3550-3650 MHz, GN Docket No. 12-354,
\emph{Report and Order and Second Notice of Proposed Rulemaking},
April 2015.

\bibitem{SASarch}
J. Ryoo, Kim, C., and Buddhikot, M. 
\emph{Design and implementation of an end-to-end architecture for 3.5 GHz shared spectrum},
IEEE International Symposium on Dynamic Spectrum Access Networks (DySPAN), 
2015.

\bibitem{GoogleSAS}
Google Access Services,
\emph{Spectrum Access System: Managing Three Tiers of Users in the 3550-3700 GHz Band}, 
available at \url{http://wireless.fcc.gov/workshops/sas_01-14-2014/panel-1/Marshall-Google.pdf}.

\bibitem{LEHR}
W. Lehr, 
\emph{Spectrum license design, Sharing and Exclusion Rights},
43rd Research Conference on Communication, Information, and Internet Policy (TPRC43), At Alexandria, VA,
Available from: William Lehr, Retrieved on: 15 March 2016.

\bibitem{WEISS}
M. B. H. Weiss, W. H. Lehr, A. Acker, M. M. Gomez,
\emph{Socio-technical considerations for Spectrum Access System (SAS) design},
IEEE International Symposium on Dynamic Spectrum Access Networks (DySPAN),
2015.

\bibitem{Incumbents}
A. Khawar, I. Ahmad and A. I. Sulyman, 
\emph{Spectrum sharing between small cells and satellites: Opportunities and challenges}, IEEE International Conference on Communication Workshop (ICCW), 
2015.

\bibitem{Radars}
M. Ghorbanzadeh, E. Visotsky, P. Moorut, W. Yang and C. Clancy,
\emph{Radar inband and out-of-band interference into LTE macro and small cell uplinks in the 3.5 GHz band},
IEEE Wireless Communications and Networking Conference (WCNC),
2015.

\bibitem{NPRM}
In the Matter of Amendment of the Commission's Rules with Regard to Commercial Operations in the 3550-3650 MHz, GN Docket No. 12-354,
\emph{Notice on Proposed Rulemaking},
Dec, 2012.

\bibitem{FNPRM}
In the Matter of Amendment of the Commission's Rules with Regard to Commercial Operations in the 3550-3650 MHz, GN Docket No. 12-354,
\emph{Further Notice of Proposed Rulemaking},
April 2014.

\bibitem{Comments_RF}
In the Matter of Amendment of the Commission's Rules with Regard to Commercial Operations in the 3550-3650 MHz, GN Docket No. 12-354,
\emph{Comments of Google Inc. on The Proposed Revised Framework},
Dec 5, 2013.


\bibitem{PCAST}
President’s Council of Advisors on Science and Technology (PCAST) Report,
\emph{Realizing the Full Potential of Government-Held Spectrum to Spur Economic Growth},
2012.





















%
%

%
%
%
%
%



%
%
\end{thebibliography}
